\documentclass[showpacs,preprintnumbers,amsmath,amssymb,pra]{revtex4}

\usepackage{amsmath}
\usepackage{amssymb}
\usepackage[dvips]{graphicx}
\usepackage[dvips]{color}
\usepackage[T2A]{fontenc}
\usepackage[cp1251]{inputenc}
\usepackage[english]{babel}

\begin{document}

\title{ On Teleportation in a System of Identical Particles}
\author{S.N.Molotkov}
\affiliation{Institute of Solid State Physics, Russian Academy of Sciences, Chernogolovka, Moscow region, 142432 Russia}
\affiliation{Faculty of Computational Mathematics and Cybernetics, M.V.Lomonosov Moscow State University, Moscow, 119992 Russia}
\author{S.P.Kulik}
\affiliation{Faculty of Physics,M.V.Lomonosov Moscow State University, Moscow, 119992 Russia} 
\author{S.S.Straupe}
\email{straups@yandex.ru}\affiliation{Faculty of Physics,M.V.Lomonosov Moscow State University, Moscow, 119992 Russia}

\begin{abstract}
The teleportation of an unknown polarization state of one of the photons in a system of identical particles has
been considered. It has been shown that spatial degrees of freedom, which are various directions of the
momentum of three photons, are of significant importance for teleportation in the system of identical particles. The inclusion of the spatial degrees of freedom increases the dimension of single-particle state space. In view of this increase, a four-dimensional subspace of two-particle states, which is similar to the state space spanned by the Bell states in a system of two distinguishable qubits, can be separated in the experimental configuration.
\end{abstract}

\maketitle

The copying (cloning) of an unknown quantum state is fundamentally forbidden \cite{WZ_Nature82}. However, the unknown quantum state can be teleported \cite{BBCJPW_PRL93}. The teleportation process was proposed in \cite{BBCJPW_PRL93} and is reduced to the transfer of the unknown quantum state from one particle to another particle of a pair in a joint entangled state. The teleportation is realized by a joint measurement of a particle in an unknown state and one particle of the Einstein–Podolsky–Rosen (EPR) pair \cite{EPR}. Then, the measurement result is reported through a classical communication channel for performing a unitary transformation of the second particle in the EPR pair, which is not measured. The EPR pair serves as a shared quantum communication channel. As a result, the second particle in the EPR pair appears exactly in the quantum state of the initial particle.

\section{Teleportation in a system of distinguishable particles}

The process of teleportation of qubits (two-level quantum systems) was initially formulated in \cite{BBCJPW_PRL93} for
distinguishable particles. Subsequent experiments on the teleportation of a state with an unknown polarization were performed with photons, which are identical particles \cite{Teleport_exp, KKS_PRL01}. It is easy to understand that teleportation in a system of three identical qubits is impossible, because the dimension of the state space of two distinguishable qubits is four, whereas the dimension of the state space of two identical qubits is three due to symmetrization. Up to date it is often stated that teleportation is impossible and that the teleportation scheme for distinguishable particles \cite{BBCJPW_PRL93} is not directly applicable to the real situation with identical particles \cite{Teleport_exp, KKS_PRL01} and, hence, the interpretation of the experimental observations \cite{Teleport_exp, KKS_PRL01} is incorrect, because it involves the results for distinguishable particles.

It was stated earlier that the identity of particles is insignificant in the teleportation process \cite{Peres}, but no clear explanation of this statement has been given as far as we know. Below, we consider the teleportation of the unknown polarization state for identical photons, which corresponds to the experimental situation in \cite{Teleport_exp, KKS_PRL01}. It will be shown that the identity of the
particles does not prevent teleportation. However, in experimental situation, in addition to the polarization degrees of freedom, the spatial degrees of freedom, which are three different directions of photon momentum, are of significant importance. As a result, the dimension of the space of two-particle states symmetrized with respect to permutations of particles is 10. The experimental procedure of
preparing the EPR pair is such that momenta of particles in the EPR pair differ in magnitude or direction. It is possible to select a four-dimensional subspace of the 10-dimensional space where the symmetric two-particle states are similar to the basis Bell states in the space of two distinguishable particles. As a result, the teleportation process is completely equivalent to the teleportation process in the system of three distinguishable particles. The only formal difference is that for distinguishable particles, the unknown state is transferred (see below) from the third particle to the second, whereas in the system of identical particles, it is impossible to indicate the particle to which the unknown quantum state is transferred. Since the particles are indistinguishable, this circumstance is of
no physical significance.

For comparison and completeness of presentation, we first present the teleportation scheme of the unknown qubit state for distinguishable particles \cite{BBCJPW_PRL93}. The entangled state of two distinguishable qubits has the form
\begin{equation}
\label{Phi+}
|\Phi^+_{12}\rangle=\frac{1}{\sqrt{2}}
(|0_1\rangle\otimes|0_2\rangle+|1_1\rangle\otimes|1_2\rangle)
\in {\cal H}_{1}\otimes{\cal H}_{2},
\end{equation}
where , $|0_{1,2}\rangle, \mbox{ } |1_{1,2}\rangle\in {\cal H}_{1,2}$ are the orthonormalized basis vectors for distinguishable particles $1$ and $2$ in corresponding two-dimensional single particle state spaces ${\cal H}_{1,2}$. Let the unknown quantum state of the third particle to be teleported have the form 
\begin{equation}\label{3d-part}
|\varphi_3\rangle=\alpha|0_3\rangle+\beta|1_3\rangle\in{\cal H}_3.
\end{equation}
The total state of three different particles is
\begin{equation}\label{total-state}
|\varphi_3\rangle\otimes |\Psi^+_{12}\rangle\in
{\cal H}_3\otimes{\cal H}_{1}\otimes{\cal H}_{2}.
\end{equation}
The teleportation process is reduced to the joint measurement of particles 1 and 3 in an entangled basis. Any quantum mechanical measurement is described by the decomposition of unity in a certain space of states, in our case, in ${\cal H}_3\otimes{\cal H}_2\otimes{\cal H}_1$,
\begin{equation}\label{1-decompos}
I_{123}=I_2\otimes I_{13}=
I_2\otimes
(
|\Psi^+_{13}\rangle\langle\Psi^+_{13}|+
|\Psi^-_{13}\rangle\langle\Psi^-_{13}|+
|\Phi^+_{13}\rangle\langle\Phi^+_{13}|+
|\Phi^-_{13}\rangle\langle\Phi^-_{13}|
)
,
\end{equation}
\begin{equation}\label{Bell-states13}
|\Phi^{\pm}_{13}\rangle=\frac{1}{\sqrt{2}}
(|0_1\rangle\otimes|0_3\rangle\pm|1_1\rangle\otimes|1_3\rangle),
\quad
|\Psi^{\pm}_{13}\rangle=\frac{1}{\sqrt{2}}
(|0_1\rangle\otimes|1_3\rangle\pm|1_1\rangle\otimes|0_3\rangle).
\end{equation}
The remarkable observation made in \cite{BBCJPW_PRL93} is that the joint state of three particles may be decomposed as
\begin{equation}
|\Psi_{123}\rangle=\frac{1}{2}
\left\{
(\alpha|0_2\rangle+\beta|1_2\rangle)|\otimes|\Phi^+_{13}\rangle+
(\alpha|0_2\rangle-\beta|1_2\rangle)|\otimes|\Phi^-_{13}\rangle+
\right.
\end{equation}
\begin{displaymath}
\left.
(\alpha|1_2\rangle+\beta|0_2\rangle)|\otimes|\Psi^+_{13}\rangle+
(\alpha|1_2\rangle-\beta|0_2\rangle)|\otimes|\Psi^-_{13}\rangle
\right\}.
\end{displaymath}
The measurement specified by Eq. (\ref{1-decompos}) of state (\ref{total-state}) gives four equiprobable outcomes corresponding to projections on two-particle entangled states. After the measurement, the state of the second particle coincides with one of the following states, depending on the outcome obtained in the measurement:
\begin{equation}\label{2d-part-psi}
|\psi^{\pm}_2\rangle=
\frac{
\mbox{Tr}_{13}\{ |\Psi_{123}\rangle\langle\Psi_{123}||\Phi^{\pm}_{13}\rangle\langle\Phi^{\pm}_{13}| \}
}
{\mbox{Tr}_{123}
\{ |\Psi_{123}\rangle\langle\Psi_{123}|(I_2\otimes|\Phi^{\pm}_{13}\rangle\langle\Phi^{\pm}_{13}|) \}
} =
\alpha|0_2\rangle\pm\beta|1_2\rangle,
\end{equation}
\begin{equation}\label{2d-part-phi}
|\phi^{\pm}_2\rangle=
\frac{
\mbox{Tr}_{13}\{ |\Psi_{123}\rangle\langle\Psi_{123}||\Psi^{\pm}_{13}\rangle\langle\Psi^{\pm}_{13}| \}
}
{\mbox{Tr}_{123}
\{ |\Psi_{123}\rangle\langle\Psi_{123}|(I_2\otimes|\Psi^{\pm}_{13}\rangle\langle\Psi^{\pm}_{13}|) \}
} =
\alpha|1_2\rangle\pm\beta|0_2\rangle,
\end{equation}
The enumeration of outcomes 1--4 requires two bits of information. Depending on the outcome of the measurements, two bits are transmitted to the receiver of the teleported state through the classical communication channel in order to perform one of the unitary rotations $I$, $\sigma_z$, $\sigma_x$, $\sigma_x\sigma_z$ for outcomes 1--4, respectively, which are independent of the initial state. After that,
the state of the second particle coincides with the initial state of particle $3$, which was in unknown state (\ref{3d-part}).

\section{Teleportation in the system of identical particles}

\begin{figure}[htb]
\begin{center}
\includegraphics[width=0.5\textwidth]{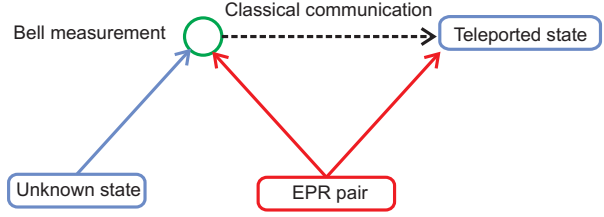}	
\end{center}
\caption{\footnotesize Scheme of teleportation process in the system of identical particles.}
\label{teleport-scheme}
\end{figure}

The scheme of the experiment with identical particles is shown on figure \ref{teleport-scheme}. 
In addition to the polarization degrees of freedom, the spatial degrees of freedom, i.e., three different momentum directions $k_1$, $k_2$ and $k_3$, are involved in the real experiment. The single-particle states are denoted as
\begin{equation}\label{single-part-sym}
|\varphi_{J}(i)\rangle, \quad  \in {\cal H},
\quad
\Omega_J=\{1H, 1V, 2H, 2V,3H,3V\}
\end{equation}
Below, the subscript $J=iH$ or $J=iV$ denotes the state with the wave vector $k_i$ and basis polarization states $H$
and $V$. Here, ${\cal H}$ is the space of single-particle states
spanned by single-particle functions (\ref{single-part-sym}).

The vector of an arbitrary multiparticle state for $n=0,1,2...$ particles can be represented in the form (see, e.g., \cite{Berezin})
\begin{equation} \label{multipart-state-sym}
|\Phi\rangle=
\left(
\begin{array}{c}
\alpha_{vac}|vac\rangle\\
\sum_{J_1} \mbox{}\alpha_{J_1}|\Phi_{J_1}\rangle\\
\sum_{J_1,J_2}\mbox{}\alpha_{J_1,J_2}|\Phi_{J_1,J_2}\rangle\\
.........\\
\sum_{J_1,J_2,..J_n}\mbox{}\alpha_{J_1,J_2,..J_n}|\Phi_{J_1,J_2,..J_n}\rangle\\
.........\\
\end{array}
\right),
\quad
\langle\Phi|\Phi\rangle=
|\alpha_{vac}|^2+\sum_{J_1}|\alpha_{J_1}|^2+...
\sum_{J_1,J_2,..J_n}|\alpha_{J_1,J_2,..J_n}|^2+..=1.
\end{equation}
Here, the component of the state vector with $n$ photons has the form
\begin{equation}\label{multipart-state-n}
|\Phi_{J_1,J_2,..J_n}\rangle=
\frac{1}{\sqrt{n!}}
\sum_{{\cal P}_n}|\varphi_{J_1}(i_1)\rangle
\otimes
|\varphi_{J_2}(i_2)\rangle
\otimes
....
|\varphi_{J_n}(i_n)\rangle,
\end{equation}
where $i_k$ in the arguments of the functions denotes the
particle number and the $n$-particle permutation is
\begin{equation}\label{permutation}
{\cal P}_n=
\left(
\begin{array}{cccc}
J_1& J_2 & ...& J_n\\
i_1&i_2&...&i_n\\
\end{array}
\right),
\quad
J_k\in \Omega_J, \quad k=0,1...n.
\end{equation}
Let the unknown single-particle quantum state have the form
\begin{equation}\label{3d-part-sym}
|\varphi_{(1)}\rangle=\alpha |\varphi_{3H}(i_3)\rangle+
\beta |\varphi_{3V}(i_3)\rangle,
\end{equation}
where subscript (1) means that the state vector refers to the single-particle space. To prepare the entangled state of the EPR pair,
parametric down-conversion is usually used in the experiment; as a result, the EPR state is obtained in the form
\begin{equation}\label{EPR-sym}
|\Phi_{(2)}^+\rangle=
\frac{1}{2}
\sum_{{\cal P}_2}
(
|\varphi_{1H}(i_1)\rangle\otimes |\varphi_{2H}(i_2)\rangle+
|\varphi_{1V}(i_1)\rangle\otimes |\varphi_{2V}(i_2)\rangle
),
\end{equation}
which is naturally symmetric with respect to permutation of particles.

The state vector for three identical particles should be symmetric with respect to permutations of all three particles, i.e.,
\begin{equation}\label{total-state-sym}
|\Psi_{(3)}\rangle=\frac{1}{2\sqrt{3}}\sum_{{\cal P}_3}|\varphi_{(1)}\rangle\otimes |\Phi_{(2)}^+\rangle =
\end{equation}
\begin{displaymath}
\frac{1}{2\sqrt{3}}
\left(
(\alpha |\varphi_{3H}(3)\rangle+\beta |\varphi_{3V}(3)\rangle)\otimes
\sum_{{\cal P}_2}
(
|\varphi_{1H}(i_1)\rangle\otimes |\varphi_{2H}(i_2)\rangle+
|\varphi_{1V}(i_1)\rangle\otimes |\varphi_{2V}(i_2)\rangle
)
\right.
+
\end{displaymath}
\begin{displaymath}
(\alpha |\varphi_{3H}(2)\rangle+\beta |\varphi_{3V}(2)\rangle)\otimes
\sum_{{\cal P}_2}
(
|\varphi_{1H}(i_1)\rangle\otimes |\varphi_{2H}(i_3)\rangle+
|\varphi_{1V}(i_1)\rangle\otimes |\varphi_{2V}(i_3)\rangle
)
+
\end{displaymath}
\begin{displaymath}
\left.
(\alpha |\varphi_{3H}(1)\rangle+\beta |\varphi_{3V}(1)\rangle)\otimes
\sum_{{\cal P}_2}
(
|\varphi_{1H}(i_3)\rangle\otimes |\varphi_{2H}(i_2)\rangle+
|\varphi_{1V}(i_3)\rangle\otimes |\varphi_{2V}(i_2)\rangle)
\right)
\end{displaymath}
One can regroup the terms in  (\ref{total-state-sym}) so that
\begin{equation}
|\Psi_{123}\rangle=
\end{equation}
\begin{displaymath}
\frac{1}{2\sqrt{3}}
\sum_{{\cal P}_3}
\left(
(\alpha |\varphi_{2H}(i_1)\rangle+\beta |\varphi_{2V}(i_1)\rangle)\otimes
(
|\varphi_{1H}(i_2)\rangle\otimes |\varphi_{3H}(i_3)\rangle+
|\varphi_{1V}(i_2)\rangle\otimes |\varphi_{3V}(i_3)\rangle
)
\right.
+
\end{displaymath}
\begin{displaymath}
(\alpha |\varphi_{2H}(i_1)\rangle-\beta |\varphi_{2V}(i_1)\rangle)\otimes
(
|\varphi_{1H}(i_2)\rangle\otimes |\varphi_{3H}(i_3)\rangle
-
|\varphi_{1V}(i_2)\rangle\otimes |\varphi_{3V}(i_3)\rangle
)
+
\end{displaymath}
\begin{displaymath}
(\alpha |\varphi_{2V}(i_1)\rangle+\beta |\varphi_{2H}(i_1)\rangle)\otimes
(
|\varphi_{1H}(i_2)\rangle\otimes |\varphi_{3V}(i_3)\rangle
+
|\varphi_{3V}(i_2)\rangle\otimes |\varphi_{1H}(i_3)\rangle)
+
\end{displaymath}
\begin{displaymath}
(\alpha |\varphi_{2V}(i_1)\rangle-\beta |\varphi_{2H}(i_1)\rangle)\otimes
(
|\varphi_{1H}(i_2)\rangle\otimes |\varphi_{3V}(i_3)\rangle
-
|\varphi_{3V}(i_2)\rangle\otimes |\varphi_{1H}(i_3)\rangle)
\end{displaymath}

In the experiment reported in \cite{KKS_PRL01}, the measurements were performed in the complete Bell basis with the use of parametric up-conversion (for details, see \cite{KKS_PRL01, Molotkov_JETPLett98}). Any quantum mechanical measurement is formally described by decomposition of unity. For identical particles, the operators in $\mbox{Sym}\{ {\cal H}^{\otimes 2}\}$, implementing the decomposition of unity, should be symmetric with respect to permutations of particles; i.e.,
\begin{equation}\label{1-decompos-sym}
I_{\mbox{Sym}\{{\cal H}^{\otimes 2}\}}=
\end{equation}
\begin{displaymath}
I^{\bot Bell}_{\mbox{Sym}\{{\cal H}^{\otimes 2}\}}+
|\Psi^+_{(2)}\rangle\langle \Psi^+_{(2)}|+
|\Psi^-_{(2)}\rangle\langle \Psi^-_{(2)}|+
|\Phi^+_{(2)}\rangle\langle \Phi^+_{(2)}|+
|\Phi^-_{(2)}\rangle\langle \Phi^-_{(2)}|,
\end{displaymath}
Here, $I^{\bot Bell}_{\mbox{Sym}\{{\cal H}^{\otimes 2}\}}$ is the identity operator in the subspace spanned by the vectors orthogonal to the Bell states
\begin{equation}\label{Phi13-sym}
|\Phi^{\pm}_{(2)}\rangle=
\frac{1}{2}
\left(
\sum_{{\cal P}_2}
|\varphi_{1H}(i_1)\rangle\otimes |\varphi_{3H}(i_2)\rangle
\pm
\sum_{{\cal P}_2}
|\varphi_{1V}(i_1)\rangle\otimes |\varphi_{3V}(i_2)\rangle
\right).
\end{equation}

\begin{equation}\label{Psi13-sym}
|\Psi^{\pm}_{(2)}\rangle=
\frac{1}{2}
\left(
\sum_{{\cal P}_2}
|\varphi_{1H}(i_1)\rangle\otimes |\varphi_{3V}(i_2)\rangle
\pm
\sum_{{\cal P}_2}
|\varphi_{1V}(i_1)\rangle\otimes |\varphi_{3H}(i_2)\rangle
\right)
.
\end{equation}
Note that the states specified by Eqs. (\ref{Phi13-sym}) and (\ref{Psi13-sym}) are the sum or difference of two terms symmetric with
respect to permutation of particles and do not change their sign under permutation of particles in contrast to the states $|\Psi^{-}_{13}\rangle$ and $|\Phi^{-}_{13}\rangle$ in Eqs. (\ref{Bell-states13}) for distinguishable particles, which are antisymmetric
with respect to the permutation of qubits 1 and 3. 

The measurements in the experiment reported in \cite{KKS_PRL01} provide outcomes only in four channels corresponding to projections on Bell states in Eq. (\ref{1-decompos-sym}).

The single-particle states after the measurements have the form
\begin{equation}\label{psi1-state-sym}
|\psi^{\pm}_{(1)}\rangle=
\frac{
\mbox{Tr}_
{\footnotesize \mbox{Sym}\{ {\cal H}^{\otimes 2}\} }
\{
|\Psi_{(3)}\rangle
\langle\Psi_{(3)}||\Phi^{\pm}_{(2)}\rangle
\langle\Phi^{\pm}_{(2)}|
\}
}
{\mbox{Tr}_
{\footnotesize \mbox{Sym}\{ {\cal H}^{\otimes 3}\} }
\{ |\Psi_{(3)}\rangle\langle\Psi_{(3)}|(|\Phi^{\pm}_{(2)}\rangle\langle\Phi^{\pm}_{(2)}|) \}
} =
\frac{1}{\sqrt{3}}
\sum_{{\cal P}_1}
(\alpha |\varphi_{2H}(i_1)\rangle\pm\beta |\varphi_{2V}(i_1)\rangle),
\end{equation}
\begin{equation}\label{phi1-state-sym}
|\phi^{\pm}_{(1)}\rangle=
\frac{
\mbox{Tr}_
{\footnotesize \mbox{Sym}\{ {\cal H}^{\otimes 2}\} }
\{
|\Psi_{(3)}\rangle
\langle\Psi_{(3)}||\Psi^{\pm}_{(2)}\rangle
\langle\Psi^{\pm}_{(2)}|
\}
}
{
\mbox{Tr}_
{\footnotesize \mbox{Sym}\{ {\cal H}^{\otimes 3}\} }
\{ |\Psi_{(3)}\rangle\langle\Psi_{(3)}|(|\Psi^{\pm}_{(2)}\rangle\langle\Psi^{\pm}_{(2)}|) \}
} =
\frac{1}{\sqrt{3}}
\sum_{{\cal P}_1}
(\alpha |\varphi_{2V}(i_1)\rangle\pm\beta |\varphi_{2H}(i_1)\rangle)
.
\end{equation}
Since the particles are indistinguishable (the particle subscript in the summand is dummy), the single-particle states given by Eqs. (\ref{psi1-state-sym})and (\ref{phi1-state-sym}) can be represented in the form

\begin{equation}\label{psi1-state-sym-2}
|\psi^{\pm}_{(1)}\rangle=
(\alpha |\varphi_{2V}(i)\rangle\pm\beta |\varphi_{2H}(i)\rangle)
\rightarrow
\alpha |0\rangle\pm\beta |1\rangle
,
\end{equation}
\begin{equation}\label{phi1-state-sym-2}
|\phi^{\pm}_{(1)}\rangle=
(\alpha |\varphi_{2H}(i)\rangle\pm\beta |\varphi_{2V}(i)\rangle)
\rightarrow
\alpha |1\rangle\pm\beta |0\rangle
.
\end{equation}
Further, depending on the outcome, similar to the case of distinguishable particles, the unitary rotation of the state occurs in Eqs. (\ref{psi1-state-sym}) and (\ref{phi1-state-sym}). After such a rotation, the teleported state coincides with the initial unknown state given by Eq. (\ref{3d-part-sym}).

\section{Conclusion}
To summarize, teleportation in the system of identical particles exists and is similar to teleportation for distinguishable particles. This is possible, because in experimental situation the spatial degrees of freedom, which are three different directions of the photon momentum, are of significant importance, in addition to the polarization degrees of freedom. As a result, the dimension of two-particle
state space symmetrized with respect to permutations of particles is 10. Only the four-dimensional subspace of the total 10-dimensional space of states is involved in the experiment; the symmetric two-particle states in this subspace are similar to the basis Bell
states in the space of two distinguishable particles. As a result, the teleportation process is physically equivalent to the teleportation process in the system of three distinguishable particles. Note again that teleportation in the system of three identical qubits disregarding the spatial degrees of freedom is impossible, because the dimension of the two-particle state space is three, rather than four, like for distinguishable qubits.

\acknowledgements{
We are grateful to M.V. Fedorov for discussion of the results. This work was supported in part by
the Russian Foundation for Basic Research, project nos. 08-02-00559 and 19-02-90036-Bel.}

\end{document}